# Identifying the 3FHL Catalog. VI. Swift Observations of 3FHL Unassociated Objects with Source Classification via Machine Learning

S. Joffre[1], R. Silver[1], M. Rajagopal[1], M. Ajello[1], N. Torres-Albà[1], A. Pizzetti[1], S. Marchesi[1,2], and A. Kaur[3]  
[1] Department of Physics and Astronomy, Clemson University, Kinard Lab of Physics, Clemson, SC 29634, USA  
[2] INAF − Osservatorio di Astrofisica e Scienza dello Spazio di Bologna, Via Piero Gobetti, 93/3, I-40129, Bologna, Italy  
[3] Department of Astronomy and Astrophysics, 525 Davey Lab, Pennsylvania State University, University Park, PA 16802, USA  
*Received 2022 July 5; revised 2022 September 16; accepted 2022 October 4; published 2022 November 29*

## Abstract

The Third Catalog of Hard Fermi Large Area Telescope Sources (3FHL) reports the detection of 1556 objects at $E > 10$ GeV. However, 177 sources remain unassociated and 23 are associated with a ROSAT X-ray detection of unknown origin. Pointed X-ray observations were conducted on 30 of these unassociated and unknown sources with Swift−XRT. A bright X-ray source counterpart was detected in 21 out of 30 fields. In five of these 21 fields, we detected more than one X-ray counterpart, totaling 26 X-ray sources analyzed. Multiwavelength data was compiled for each X-ray source detected. We find that 21 out of the 26 X-ray sources detected display the multiwavelength properties of blazars, while one X-ray source displays the characteristics of a Galactic source. Using trained decision tree, random forest, and support vector machine models, we predict all 21 blazar counterpart candidates to be BL Lacertae objects (BL Lacs). This is in agreement with BL Lacs being the most populous source class in the 3FHL.

*Unified Astronomy Thesaurus concepts:* Active galactic nuclei (16); Blazars (164); Gamma-ray astronomy (628); Random Forests (1935); Support vector machine (1936); X-ray identification (1817)

*Supporting material:* interactive figure

## 1. Introduction

Active galactic nuclei (AGN) are luminous galactic centers that are believed to host accreting supermassive black holes ($M_{BH} \gtrsim 10^6 M_\odot$; e.g., Urry & Padovani 1995; Risaliti & Elvis 2004; Paliya et al. 2016; Acero et al. 2016; Marcotulli et al. 2017). Blazars are AGN with variable emission that is likely caused by a relativistic jet directed toward the observer (viewing angle $\theta_v < 10°$; Blandford & Rees 1978). Moreover, blazar emission spans the entire electromagnetic spectrum (Böttcher 2007). Blazars are currently categorized into two classes by their optical emission spectrum: flat-spectrum radio quasars (FSRQs) and BL Lacertae objects (BL Lacs/BLLs). FSRQs are distinct from BL Lacs, as their optical spectra exhibit emission lines with a minimum width of 5 Å (Urry & Padovani 1995; Ajello et al. 2014). Notably, blazars are one of the most prevalent astrophysical emitters in the high-energy gamma-ray band ($E > 100$ MeV). Indeed, blazar emission accounts for ∼50% of the extragalactic gamma-ray background (EGB; Ajello et al. 2015; Marcotulli et al. 2020). Blazars are scientifically valuable because they provide a natural, lighthouse-like probe into the extragalactic background light (EBL), the sum of all light ever produced in the universe (LAT Collaboration 2018). At $E > 10$ GeV, the interaction of photons with the EBL ($\gamma\gamma \to e^-e^+$) attenuates gamma-ray emission as a function of redshift (LAT Collaboration 2018). For this reason, blazars have been utilized to constrain the Hubble constant (Domínguez et al. 2019) and constrain the gamma-ray horizon (Ajello et al. 2015; Ackermann et al. 2016).

Gamma-ray surveys are necessary to study the high-energy emission from blazars, as well as other gamma-ray sources, such as pulsars and gamma-ray bursts. The Energy Gamma-Ray Experiment Telescope (EGRET; Thompson et al. 2005) provided a full sky measurement above 30 MeV. Its successor, the Fermi satellite, is equipped with next-generation instruments for gamma-ray astronomy, namely the Large Area Telescope (LAT; Atwood et al. 2009) and the Gamma-Ray Burst monitor (GRB; Meegan et al. 2009).

The sensitivity and angular resolution of the LAT are respective factors of 100 and 3 better than those of EGRET. LAT observations led to the creation of multiple broadband catalogs, such as the 1FHL (Ackermann et al. 2013), 2FHL (Ackermann et al. 2016), 3FGL (LAT Collaboration 2015), and 4FGL (Abdollahi et al. 2020) catalogs. Using the first seven years of LAT data, the Third Catalog of Hard Fermi Large Area Telescope Sources (3FHL; Ajello et al. 2017) reports the detection of 1556 sources with energies between 10 GeV and 2 TeV. This provides invaluable data for investigating blazars.

At the time of publication, 177 objects within the 3FHL catalog were unassociated and 23 were reported as having an association with a ROSAT detection of unknown origin. These 23 associations are classified as unknown. Completing this catalog is vital to fully describing the numerous source classes that emit gamma-rays and the emission mechanisms that underlie each one. Notably, the associated 3FHL sources are primarily extragalactic (80%) with BL Lacs accounting for 61% of the extragalactic population (Ajello et al. 2017).

Furthermore, a completed 3FHL, (i.e., with source redshifts and class identifications) may provide a useful data set to plan observations for the next generation of ground-based, gamma-ray observatories such as the Cerenkov Telescope Array (CTA; Hassan et al. 2017).

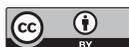

A major challenge of identifying blazars, such as BL Lacs, is the large positional uncertainty in gamma-ray detectors. Specifically, $2'.3$ is the 3FHL's median error radius at a 95% confidence level. This large uncertainty makes identifying a counterpart a challenging task. Therefore, follow-up observations at lower energies are needed to identify the counterparts to the gamma-ray sources.

X-ray follow-up campaigns can be very successful in locating gamma-ray counterparts, with a precision of a few arcseconds (see Stroh & Falcone 2013; Paiano et al. 2017b; Silver et al. 2020; Kerby et al. 2021a). The X-ray Telescope (XRT) on board the Neil Gehrels Swift Observatory (Gehrels 2004) has a positional accuracy of $3''$ and is sensitive in the energy range of 0.2–10 keV (Burrows et al. 2005). Using Swift–XRT, Kaur et al. (2019) provided a likely association for 52 3FHL sources, while Silver et al. (2020) provided an additional 22. This work follows the procedure of these previous campaigns, analyzing Swift observations for 30 3FHL unassociated or unknown sources. Doing so, we seek to identify likely counterpart associations and better understand the nature of these unknown sources.

We also analyze data from the Swift UltraViolet/Optical Telescope (UVOT; Roming et al. 2005), whose precise location of these blazar candidates is critical for optical follow-ups by ground-based observatories. Ground-based observations could enable redshift measurements for these sources (Marchesi et al. 2018; Desai et al. 2019; Rajagopal et al. 2021).

In addition, machine-learning algorithms such as decision trees (DT; Breiman 2001), random forests (RF; Breiman 2001), and support vector machines (SVM; Zhou 2021) have been trained to classify blazar candidates as either BL Lacs or FSRQs. Classifications from the machine-learning models are based on spectral properties of the candidates, such as color differences and spectral photon index.

The organization of this paper is as follows: Section 2 outlines the X-ray and UV/optical data analysis, while Section 3 analyzes the results of the 30 unassociated 3FHL sources that were studied. Section 4 describes the classification of the sources with the machine-learning models. Section 5 summarizes the results of this investigation.

## 2. Data

Of the 1556 sources detected in the 3FHL, originally, 177 were unassociated and 23 sources had an unknown classification. Of the combined 200 (unassociated or unknown classification), so far 74 have been identified with an X-ray counterpart (Kaur et al. 2019; Silver et al. 2020). Following this procedure, we proposed for Swift–XRT to observe 20 unobserved, unassociated sources for 4 ks each (Proposal 1720095, PI: Ajello). In the spring of 2022, Swift was placed on safe mode, which excluded three sources from observation. In addition, we found archival Swift observations (with exposures of at least 2 ks) for an additional 13 unassociated 3FHL objects. For all data processing, HEASoft version 6.28[4] was used (NASA High Energy Astrophysics Science Archive Research Center (HEASARC) 2014).

### 2.1. Swift Data Analysis

#### 2.1.1. XRT

For the 30 unassociated 3FHL objects, if multiple observations were taken, the event files were stacked into a single summed event file following XSELECT's user guide.[5] Similarly, images were stacked in compliance with XIMAGE's instruction manual.[6] The XRT instrument software used in analysis was version 3.6.0 (2020 June 26) and calibrated with the most recent CALDB version at the time of analysis (1.0.2).

Sources were detected above the $5\sigma$ threshold using the `detect` command in XIMAGE (version 4.5.1). X-ray sources are labeled as candidate counterparts if they fall within the 95% positional uncertainty region of the 3FHL source or within the 95% uncertainty region of the Fermi Large Area Telescope Fourth Source Catalog, Data Release 2 sources (4FGL-DR2; Abdollahi et al. 2020; Ballet et al. 2020) associated with a given 3FHL source. Exceptions were made for promising candidates that lay outside of the 3FHL and 4FGL uncertainty regions (average distance of $1'.6$; see Table 1 and Sections 3.1.1 −3.1.5). The precise source position was extracted with the `xrtcentroid` tool.

To extract the source spectra, a $10''$–$15''$ region centered on the counterpart candidate was used. The background region chosen was an annulus, centered on the source, with an inner radius and an outer radius of $35''$ and $71''$, respectively. These values guaranteed no overlap between the background and source regions.

The X-ray spectra were rebinned to have at least one count per bin, except the bright detection in 3FHL J0121.8+3808, whose spectrum was binned at ten counts per bin. All spectral analysis was done in XSPEC[7] (Arnaud 1996; version 12.11.1). For proper statistical analysis of the fit, the low-count statistics required the use of the Cash statistic (C-statistic; Cash 1979), while we used $\chi^2$ statistics for 3FHL J0121.8+3808. The Galactic column densities in the direction of the source were taken from Kalberla et al. (2005) and Bekhti et al. (2016), with their absorption modeled using the Tuebingen–Boulder model (Wilms et al. 2000). The $0.3-10$ keV spectra was modeled with a power law. No source spectra exhibited any additional features. The X-ray source parameters are reported in Table 1.

#### 2.1.2. UVOT

UVOT has six broadband filters, which span from 1928 to 5468 Å. After an X-ray counterpart was detected, the UVOT images of the source were stacked following the steps outlined on the UK Swift Science Centre's UVOT Data Analysis webpage.[8] Using the X-ray counterpart coordinates, the source emission was extracted using a circular region with a radius of 5".

In 14 (out of 22) cases, a UVOT source was found within the uncertainty region (median radius of $4''.13$) of the detected X-ray source. If no source was detected with UVOT, a $3\sigma$ upper limit was computed at the position of the X-ray source. Due to the high density of UV/optical sources, a $20''$ circular background region was placed near the target in order to avoid contamination from surrounding sources. Both the AB magnitude and the corresponding positional uncertainty of the UVOT sources were extracted with `uvotsource`.

Each magnitude was corrected for Galactic extinction along the line of sight to the source following the results in Table 5 of

---

[4] https://heasarc.gsfc.nasa.gov/docs/software.html
[5] https://hera.gsfc.nasa.gov/docs/software/heasoft/ftools/xselect/xselect.html
[6] https://heasarc.gsfc.nasa.gov/docs/xanadu/ximage/manual/ximage.html
[7] https://heasarc.gsfc.nasa.gov/xanadu/xspec/
[8] https://www.swift.ac.uk/analysis/uvot/image.php

Table 1
Swift−XRT Analysis

| 3FHL Name | XRT Name[b] | X-Ray R.A (HH:MM:SS.SS) | X-Ray Decl. (°:′:″) | Exp. Time (ks) | S/N[c] | $\Gamma_X$[d] | $N_H$[e] ($\times 10^{22}$ cm$^{-2}$) | Flux[f] (cgs) | RL[g] | C-stat/dof | In 95%[j] |
|---|---|---|---|---|---|---|---|---|---|---|---|
| **3FHL J0121.8+3808 − 1**[h] | J012206+380445 | 01:22:06.20 | 38:04:45.40 | 5.9 | 54 | 2.50 ± 0.11 | 0.054 | $0.30^{+0.01}_{-0.01}$ | ... | 1.11[i] | No, 2.07′ |
| **3FHL J0121.8+3808 − 2**[h] | J012219+380801 | 01:22:19.34 | 38:08:00.98 | 5.9 | 23 | 1.86 ± 0.26 | 0.055 | $0.054^{+0.01}_{-0.01}$ | ... | 1.20 | No, 3.3′ |
| 3FHL J0221.4+2512 | J022135+251418 | 2:21:26.97 | 25:14:33.67 | 20.3 | 5 | 2.00[a] | 0.0581 | $0.056^{+0.03}_{-0.02}$ | ... | 1.11 | Yes |
| **3FHL J0233.0+3742 − 1**[h] | J023308+374158 | 02:33:07.76 | 37:41:57.69 | 4.2 | 6 | 2.12 ± 0.72 | 0.0518 | $0.35^{+0.16}_{-0.12}$ | 152 ± 42 | 1.10 | Yes |
| **3FHL J0233.0+3742 − 2**[h] | J023256+373843 | 02:32:55.80 | 37:38:43.41 | 4.2 | 19 | 1.87 ± 0.21 | 0.0526 | $2.22^{0.31}_{-0.28}$ | 76[k] ± 3 | 0.68 | No, 0.75′ |
| **3FHL J0319.2−7045 − 1**[h] | J032009−704535 | 03:20:09.38 | −70:45:34.58 | 6.2 | 46 | 1.73 ± 0.09 | 0.0514 | $7.71^{+0.47}_{-0.45}$ | 262 ± 30 | 0.93 | No, 1.3′ |
| **3FHL J0319.2−7045 − 2**[h] | J032007−704320 | 03:20:07.32 | −70:43:20.10 | 6.2 | 8 | 1.76 ± 0.63 | 0.0511 | $0.26^{+0.10}_{-0.08}$ | ... | 0.82 | No, 1.5′ |
| 3FHL J0402.9+6433 | J040255+643510 | 04:02:55.08 | 64:35:09.61 | 3.6 | 6 | 3.00 ± 1.30 | 0.251 | $0.52^{0.30}_{-0.22}$ | 20[k] ± 1 | 1.65 | Yes |
| **3FHL J0459.3+1921** | J045928+192213 | 04:59:27.61 | 19:22:13.40 | 3.8 | 13 | 2.05 ± 0.34 | 0.2040 | $1.93^{+0.39}_{-0.34}$ | 7[k] ± 1 | 1.16 | Yes |
| **3FHL J0500.6+1903** | J050043+190315 | 05:00:42.98 | 19:03:14.80 | 11.2 | 33 | 1.98 ± 0.14 | 0.1920 | $2.99^{0.25}_{-0.24}$ | ... | 0.72 | Yes |
| 3FHL J0501.0+2425 | J050107+242316 | 05:01:06.98 | 24:23:16.32 | 4.1 | 10 | 2.68 ± 0.63 | 0.2440 | $0.94^{0.30}_{-0.25}$ | 37 ± 3 | 0.85 | Yes |
| **3FHL J0838.5+4006** | J083903+401547 | 08:39:02.93 | 40:15:47.29 | 3.8 | 8 | 2.09 ± 0.61 | 0.034 | $0.46^{+0.17}_{-0.13}$ | ... | 0.87 | No, 0.7′ |
| **3FHL J0901.5+6712** | J090135+671318 | 09:01:34.68 | 67:13:17.79 | 6.2 | 8 | 2.47 ± 0.61 | 0.0459 | $0.66^{0.15}_{-0.18}$ | 40 ± 3 | 0.61 | Yes |
| **3FHL J0950.6+6357** | J095038+635957 | 09:50:38.14 | 63:59:56.82 | 3.7 | 4 | 2.84 ± 1.03 | 0.0448 | $0.21^{0.13}_{-0.09}$ | 12[k] ± 2 | 2.12 | Yes |
| **3FHL J1127.8+3615 − 1**[h] | J112759+362033 | 11:27:59.08 | 36:20:33.09 | 4.2 | 5 | 2.24 ± 1.01 | 0.0222 | $0.29^{+0.16}_{-0.11}$ | 2577 ± 108 | 0.69 | No, 1.1′ |
| **3FHL J1127.8+3615 − 2**[h] | J112741+362051 | 11:27:40.74 | 36:20:50.85 | 4.2 | 6 | 2.33 ± 0.95 | 0.0233 | $0.20^{+0.13}_{-0.09}$ | ... | 1.24 | No, 1.7′ |
| **3FHL J1421.5−1654** | J142129−165455 | 14:21:29.22 | −16:54:55.13 | 2.1 | 10 | 2.09 ± 0.32 | 0.0666 | $2.16^{+0.45}_{-0.39}$ | 23 ± 2 | 0.81 | Yes |
| 3FHL J1626.3−4915 − 1[h] | J162703−491231 | 16:27:02.67 | −49:12:30.73 | 19.2 | 10 | −0.32 ± 0.88 | 1.98 | $2.63^{+0.80}_{-0.67}$ | ... | 1.07 | No, 2.4′ |
| 3FHL J1626.3−4915 − 2[h] | J162609−491746 | 16:26:08.77 | −49:17:45.70 | 19.2 | 7 | 4.41 ± 1.07 | 2.00 | $6.97^{+2.68}_{-2.14}$ | ... | 0.72 | Yes |
| 3FHL J1729.9−4148 | J172947−414828 | 17:29:46.64 | −41:48:28.17 | 5.7 | 7 | 1.63 ± 1.10 | 0.326 | $0.37^{+0.20}_{-0.15}$ | ... | 1.28 | Yes |
| **3FHL J1808.7+2420** | J180846+241906 | 18:08:45.67 | 24:19:06.22 | 5.2 | 19 | 1.72 ± 0.22 | 0.09 | $1.98^{+0.29}_{-0.27}$ | 82 ± 11 | 0.67 | Yes |
| **3FHL J1849.3−6448** | J184926−644929 | 18:49:26.26 | −64:49:28.98 | 3.7 | 14 | 2.32 ± 0.23 | 0.0516 | $2.71^{+0.38}_{-0.35}$ | ... | 1.03 | Yes |
| 3FHL J2026.7+3449 | J202638+345022 | 20:26:38.49 | 34:50:22.18 | 6.7 | 16 | 1.68 ± 0.33 | 0.654 | $2.45^{+0.42}_{-0.37}$ | ... | 0.99 | Yes |
| 3FHL J2109.6+3954 | J210936+395511 | 21:09:36.35 | 39:55:11.53 | 3.3 | 8 | 3.60 ± 0.81 | 0.294 | $1.59^{+0.12}_{-0.09}$ | ... | 0.65 | Yes |
| **3FHL J2317.8+2839** | J231740+283955 | 23:17:40.39 | 28:39:55.49 | 14.1 | 7 | 2.96 ± 0.78 | 0.0666 | $0.12^{+0.05}_{-0.04}$ | 89[k] ± 4 | 0.54 | Yes |
| 3FHL J2358.4−1808 | J235837−180718 | 23:58:36.78 | −18:07:18.49 | 6.0 | 27 | 2.51 ± 0.17 | 0.0228 | $2.34^{+0.24}_{-0.23}$ | 74 ± 6 | 0.92 | Yes |

**Notes.** Table 1 outlines the X-ray sources that were detected as possible blazar counterpart candidates in the 30 fields analyzed. The corresponding source information, fitting parameters, and results are reported here. Bolded sources indicate high latitude ($|b| > 10°$).
[a] The model fit did not yield well-constrained results. Therefore, the photon index was frozen at 2.00.
[b] Name for the sources detected by XRT. All sources detected by XRT are prefaced with the SWIFT designation (e.g., SWIFT J023308+374158).
[c] Signal-to-noise ratio.
[d] X-ray photon index.
[e] Galactic column density.
[f] Unabsorbed flux in the 0.3−10 keV band ($\times 10^{-12}$ erg cm$^{-2}$ s$^{-1}$).
[g] Radio loudness, i.e., ratio of flux density at 5GHz and flux density in the B band.
[h] Last digit was added to distinguish between the multiple sources present in a single Swift−XRT field.
[i] This value was calculated with $X^2$/dof instead of C-stat/dof.
[j] Distance from boundary of 95% uncertainty region to counterpart is reported if the counterpart candidate was not within the 95% uncertainty region.
[k] Radio sources that correspond to the VLASS1.1 catalog and therefore require a 15% correction as recommended by https://science.nrao.edu/vlass/data-access/vlass-epoch-1-quick-look-users-guide.

Table 2
Swift-UVOT Magnitudes

| Source Name | W2 | M2 | W1 | U | B | V |
|---|---|---|---|---|---|---|
| **3FHL J0121.8+3808 − 1**[a] | 19.26±0.04 | 20.50 ± 0.04 | 20.34 ± 0.04 | 20.24 ± 0.04 | 20.01 ± 0.07 | 20.07 ± 0.12 |
| **3FHL J0221.4+2512** | 22.24 ± 0.26 | 23.5 ± 0.21 | 20.12 ± 0.08 | 20.78 ± 0.11 | ... | ... |
| **3FHL J0233.0+3742 − 1**[a] | >21.51 | 21.28 ± 0.25 | 20.77 ± 0.26 | 19.87 ± 0.21 | 18.13 ± 0.09 | 17.00 ± 0.08 |
| **3FHL J0233.0+3742 − 2**[a] | 21.02 ± 0.16 | 21.02 ± 0.21 | 20.43 ± 0.19 | 20.01 ± 0.23 | 19.91 ± 0.27 | >18.90128 |
| **3FHL J0319.2−7045 − 1**[a] | 20.68 ± 0.13 | 20.63 ± 0.15 | 20.69 ± 0.13 | 20.14 ± 0.14 | 19.99 ± 0.24 | 19.28 ± 0.18 |
| 3FHL J0402.9+6433 | >17.03 | >17.59 | >17.15 | >17.32 | >17.08 | >16.94 |
| 3FHL J0459.3+1921 | >18.59 | >18.87 | >18.46 | >18.16 | >17.68 | >17.22 |
| 3FHL J0500.6+1903 | >19.12 | >19.33 | >18.86 | >18.75 | >18.26 | >17.84 |
| 3FHL J0501.0+2425 | >18.54 | >18.79 | >18.24 | >18.16 | >17.69 | >17.25 |
| **3FHL J0838.5+4006** | 21.17 ± 0.12 | ... | 20.87 ± 0.10 | ... | ... | ... |
| **3FHL J0901.5+6712** | >21.77 | >20.78 | >20.66 | >20.11 | >18.62 | >18.93 |
| **3FHL J0950.6+6357** | > 21.98 | >21.78 | >20.91 | >20.17 | >19.47 | >18.84 |
| **3FHL J1127.8+3615 − 1**[a] | 20.99 ± 0.13 | 20.62 ± 0.14 | 20.62 ± 0.19 | 20.05 ± 0.25 | >19.71 | >18.90 |
| **3FHL J1421.5−1654** | 20.79 ± 0.28 | 19.97 ± 0.21 | >20.47 | >19.72 | >19.11 | >17.95 |
| 3FHL J1729.9−4148 | >15.51 | >16.21 | >14.65 | 15.99 ± 0.14 | ... | ... |
| **3FHL J1808.7+2420** | 20.59 ± 0.13 | 20.52 ± 0.16 | 20.37 ± 0.18 | 20.09 ± 0.21 | >19.97 | 18.91 ± 0.28 |
| **3FHL J1849.3−6448** | 20.18 ± 0.11 | 20.05 ± 0.11 | 19.90 ± 0.14 | 19.45 ± 0.13 | ... | 18.12 ± 0.16 |
| **3FHL J2109.6+3954** | >18.21 | ... | 18.66 ± 0.26 | ... | ... | ... |
| **3FHL J2358.4−1808** | 18.65 ± 0.04 | 18.88 ± 0.10 | 18.73 ± 0.05 | 18.01 ± 0.04 | 18.94 ± 0.14 | 18.29 ± 0.29 |
| **3FHL J2317.8+2839** | 20.99 ± 0.10 | 20.86 ± 0.10 | 20.57 ± 0.14 | 20.38 ± 0.26 | >20.08 | >19.14 |

**Notes.** Table 2 outlines AB magnitudes for all counterpart candidates that lie at $|b| > 5°$. Bolded sources indicated high latitude ($|b| > 10°$).
[a] The additional digit was used to distinguish sources where more than one was present in a single field.

Kataoka et al. (2008). The reddening factor was calculated using NASA/IPAC's Galactic Dust Reddening and Extinction online tool[9] (Schlafly & Finkbeiner 2011; IRSA 2022). This process was sequentially repeated for each band in UVOT. This calculation, although reliable for sources outside of the Galactic plane, is limited because it performs accurately only if characterized by a single dust temperature.[10] Therefore, this characterization is not reliable near the Galactic plane, due to a high probability of varied dust temperatures. SWIFT J162703−491231, SWIFT J162609−491746, and SWIFT J202638+345022 all have Galactic latitudes $|b| < 2°$, and consequently their AB-corrected magnitudes were not reported. The final corrected AB magnitude values for each available UVOT filter are reported in Table 2.

### 2.2. Archival Data

Additional archival data at radio and infrared (IR) wavelengths of the candidate counterparts were obtained from SIMBAD[11] (Wenger et al. 2000) and VizieR (Ochsenbein et al. 2000) searching within 5″ from the best-fit X-ray position. The crossmatch service provided by CDS, Strasbourg[12] was also used. These are listed in Appendix A.

### 3. Results

Thus far, ∼77% of the classified 3FHL sources have been associated with blazars (BL Lacs, FSRQS, or blazars of uncertain types, abbreviated BCUs). Of the 1235 high-latitude ($|b| > 10°$) sources in the 3FHL, 1080 (or ∼90%) of them are blazars. Of the 21 3FHL sources reported in Table 1, 16 of them are high-latitude.

With regards to Figure 1, the unassociated sources studied have a 3FHL photon index range of 1.44 to 3.84, and a gamma-ray flux in between $4.7 \times 10^{-13}$ erg cm$^{-2}$ s$^{-1}$ and $4.8 \times 10^{-12}$ erg cm$^{-2}$ s$^{-1}$, omitting 3FHL J1626.3−4915, the source suspected to be of Galactic origin (see Section 3.4). Of the classified 3FHL sources that fall within this gamma-ray photon index and flux range, 95% are known blazars. At high latitudes, this fraction becomes 97%. We therefore expect most of our sources to be blazars.

Blazars also represent 78% of the classified low-latitude sources ($|b| < 10°$) within the same range of photon indices and fluxes. Therefore, we expect four of our five low-latitude sources to be blazars.

Of the 30 unassociated and unknown 3FHL sources analyzed, we identified an X-ray counterpart within or near (∼3′) the 3FHL 95% uncertainty region for 21 sources. Of those 21 3FHL regions, 16 had a single X-ray counterpart while five of them had two or more X-ray sources detected. The counterpart selection for the double X-ray detections are discussed in detail in Sections 3.1.1–3.1.5.

The corresponding Swift−XRT flux (0.3–10 keV) versus photon index plot as shown in Figure 2 demonstrates the difficulty in using these features to disentangle the differences between Galactic and extragalactic sources. This lack of clear distinction between the groups led us to use other multiwavelength properties to determine the likely classification for our unassociated sources (as done by Massaro et al. 2012, Kaur et al. 2019, and Silver et al. 2020).

One such feature is the Wide-field Survey Explorer (WISE; Cutri et al. 2021) blazar strip in the infrared (IR) color–color space identified by Massaro et al. (2012); see their Figures 1 and 2. As Figure 3 indicates, BL Lacs and FSRQs inhabit distinct positions in this space. BL Lacs occupy space that is bluer than their redder FSRQ neighbors. Also, this color space demonstrates that blazars tend to have greater magnitudes (lower fluxes) in W1 (3.4 μm) than W2 (4.6 μm), while Galactic sources demonstrate the reverse. Many of our sources

---
[9] https://irsa.ipac.caltech.edu/applications/DUST/
[10] See cautionary notes here: https://irsa.ipac.caltech.edu/applications/DUST/docs/background.html
[11] http://simbad.u-strasbg.fr/simbad/
[12] http://cdsxmatch.u-strasbg.fr/xmatch

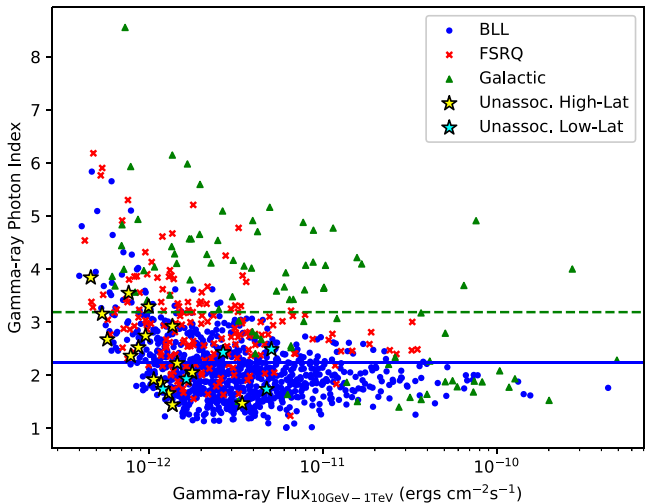

**Figure 1.** Gamma-ray photon index vs. gamma-ray flux (10 GeV–1 TeV) from the 3FHL catalog. Blue points and red Xs represent the distribution of known blazar types (BL Lac and FSRQ, respectively). The green triangles correspond to Galactic sources. The stars denote our unassociated sources, most of which fall in regions dominated by known blazars. The average known blazar photon index is designated by the blue solid line. The green dashed line is the average photon index of Galactic sources.

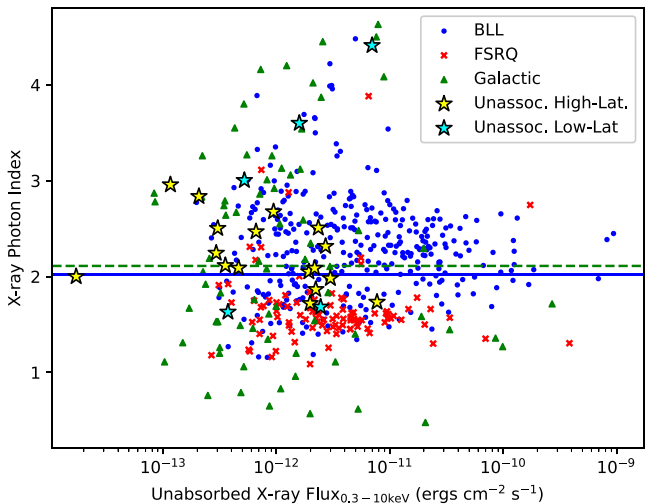

**Figure 2.** Plot of X-ray photon index vs. X-ray flux (0.3–10 keV) for 3FHL known blazars and Galactic sources. The blue line indicates the average photon index of known blazars, while the green dashed line is for Galactic sources. Stars designate the unassociated sources analyzed in this work.

have infrared colors compatible with those of blazars (specifically BL Lacs).

The single source with a very low W2−W3 color corresponds to SWIFT J162609−491746 and has a Galactic latitude of −0°.11. The colors and low latitude make this source more compatible with the characteristics of a Galactic source. This is further discussed in Section 3.4.

Table 3 reports multiwavelength (radio, IR, and UV) characteristics of the identified X-ray counterparts. Moreover, Tables 4 and 5 show that 10 sources are spatially compatible with known blazars.

### 3.1. Fields with Multiple X-Ray Sources

Of the 30 fields analyzed, five had multiple X-ray sources in or near the 95% positional uncertainty region. We used the following criteria for selecting the most likely X-ray counterpart in a given field:

1. Position falls within the 3FHL/4FGL 95% uncertainty radius
2. Exhibits a power-law X-ray spectrum
3. Has WISE IR colors consistent with those of blazars
4. Position consistent with that of a radio source.

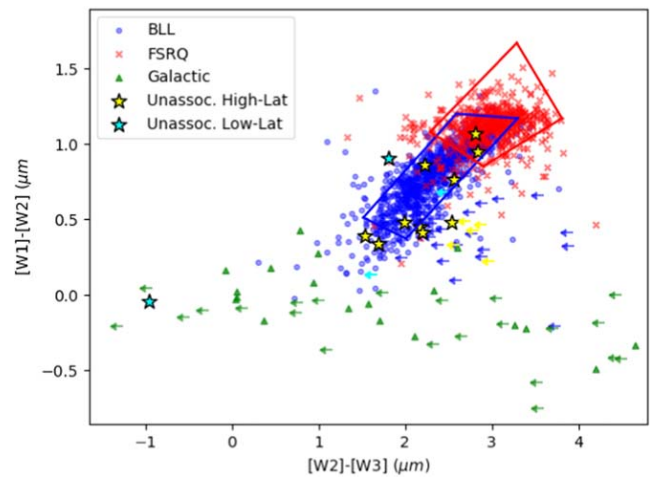

**Figure 3.** WISE blazar strip in the color space W1 (3.4 μm)−W2 (4.6 μm) vs. W2–W3 (12 μm). The red and blue regions outlined by polygons (as reported in Massaro et al. 2012) represent the FSRQ and BLL regions, respectively. Arrows represent an upper limit. Note that not all sources analyzed are represented in this plot, due to lack of data in one (or both) WISE colors. For additional clarity, an interactive figure is available in the online journal, or it can be accessed here: https://authortools.aas.org/AAS40568/figNint.html. The interactive plot has multiple data layers that can be toggled on or off by selecting the checkboxes below the plot. Above the plot is a toolbar where one can select the ability to click and drag around on the plot, as well as zoom in with either a scrolling feature or by dragging a box around the area of interest.

#### 3.1.1. 3FHL J0121.8+3808

Both SWIFT J012206+380445 and SWIFT J012219 +380801 (see Figure 4) lie distinctly outside of the 3FHL region. Both lack coincidence with a radio counterpart. It is possible that additional time on Swift−XRT may be necessary to distinguish a blazar candidate inside the 3FHL region. Given that SWIFT J012206+380445 has WISE IR colors compatible with blazars while SWIFT J012219+380801 does not, we believe SWIFT J012206+380445 is the most likely counterpart to 3FHL J0121.8+3808.

#### 3.1.2. 3FHL J0319.2−7045

3FHL J0319.2−7045 has two sources that lie outside the 3FHL and 4FGL 95% uncertainty regions: SWIFT J032009 −704535 and SWIFT J032007−704320 (see Figure 5). While SWIFT J032009−704535 was significantly brighter in the 0.3 −10.0 keV band (7.71 versus $0.26 \times 10^{-12}$ erg cm$^{-2}$ s$^{-1}$), both were initially analyzed due to their close proximity to the 3FHL 95% uncertainty region (∼1.5′). SWIFT J032009−704535 (the lower source in Figure 5) had WISE IR colors compatible with blazars. Additionally, SWIFT J032009−704535 is coincident with a radio source. Furthermore, SWIFT J032009−704535 is positionally consistent with MRSS 054−102986, the BCU associated to the identified 4FGL counterpart. Meanwhile, SWIFT J032007−704320 lacked the typical characteristics of a

Table 3
Multiwavelength Data

| 3FHL Name | XRT Name[a] | XRT Unc. (″) | Radio | 2MASS | AllWISE | Ultraviolet |
|---|---|---|---|---|---|---|
| 3FHL J0121.8+3808 − 1[b] | J012206+380445 | 3.57 | ILT J012205.88+380445.2 | J01220586+3804457 | J012205.86+380445.8 | GALEX J012205.8+380445 |
| 3FHL J0221.4+2512 | ... | ... | NVSS 022126+251436 | J02212698+2514338 | J022126.96+251433.6 | ... |
| 3FHL J0233.0+3742 − 1[b] | J023308+374158 | 4.55 | NVSS J023308+374201 | J02330797+3741597 | J023307.99+374159.6 | ... |
| 3FHL J0233.0+3742 − 2[b] | J023256+373843 | 3.72 | VLASS1 J023256.08+373846.3 | J02325605+3738464 | J023256.09+373846.2 | GALEX J023307.8+374200 |
| 3FHL J0319.2−7045 − 1[b] | J032009−704535 | 3.55 | SUMSS J032009.3−704535 | ... | J032009.21−704533.6 | GALEX J032009.1−704533 |
| 3FHL J0402.9+6433 | J040255+643510 | 5.02 | NVSS J040254+643509 | J04025445+6435101 | J040254.43+643510.0 | ... |
| 3FHL J0459.3+1921 | J045928+192213 | 3.90 | VLASS1 J045927.50+192215.3 | ... | J045927.49+192214.9 | ... |
| 3FHL J0500.6+1903 | J050043+190315 | 3.58 | NVSS 050043+190310 | ... | ... | ... |
| 3FHL J0501.0+2425 | J050107+242316 | 4.10 | NVSS J050106+242316 | J05010691+2423183 | J050106.90+242317.6 | ... |
| 3FHL J0838.5+4006 | J083903+401547 | 4.32 | NVSS 083903+401546 | J08390308+4015455 | J083903.09+401545.6 | GALEX J083903.1+401545 |
| 3FHL J0901.5+6712 | J090135+671318 | 4.46 | NVSS J090133+671317 | ... | ... | GALEX J090134.1+671316 |
| 3FHL J0950.6+6357 | J095038+635957 | 5.53 | VLASS1 J095037.96+635957.3 | ... | J095037.97+635957.7 | ... |
| 3FHL J1127.8+3615 − 1[b] | J112759+362033 | 4.88 | NVSS J112758+362028 | ... | J112758.88+362028.4 | GALEX J112759.2+362028 |
| 3FHL J1421.5−1654 | J142129−165455 | 4.20 | VLASS1 J142128.99−165455.7 | ... | J142128.94−165455.4 | GALEX J142129.0−165457 |
| 3FHL J1626.3−4915 − 2[b] | J162609−491746 | 4.13 | ... | ... | J162608.74−491744.1 | ... |
| 3FHL J1729.9−4148 | J172947−414828 | 4.15 | PMN J1729−4148 | J17294637−4148283 | J172946.47−414828.6 | ... |
| 3FHL J1808.7+2420 | J180846+241906 | 3.72 | NVSS 180845+241907 | ... | J180845.69+241905.7 | ... |
| 3FHL J1849.3−6448 | J184926−644929 | 3.90 | ... | ... | J184926.42−644930.7 | ... |
| 3FHL J2026.7+3449 | J202638+345022 | 3.78 | ... | ... | ... | IGAPS J202638.61+345024.3 |
| 3FHL J2109.6+3954 | J210936+395511 | 4.32 | VLASS1 J210936.18+395513.5 | ... | J210936.14+395513.5 | ... |
| 3FHL J2317.8+2839 | J231740+283955 | 4.17 | VLASS1 J231740.21+283955.8 | ... | J231740.22+283955.9 | GALEX J231740.1+283956 |
| 3FHL J2358.4−1808 | J235837−180718 | 3.62 | NVSS 235836−180718 | J23583682−1807175 | J235836.83−180717.4 | GALEX 2668886394254922435 |

**Notes.** Table 3 outlines the spatial coincidence of multiwavelength observations with the X-ray detection made by Swift−XRT.
[a] Sources detected with Swift−XRT referenced in this paper are prefaced with the SWIFT designation (e.g., SWIFT J023308+374158).
[b] The additional digit was used to distinguish sources where more than one potential X-ray counterpart was present in a single field.

Table 4
Spatial Coincidence between 4FGL-DR2 and SWIFT-reported Counterparts

| 4FGL Name | 4FGL Class | Counterpart R.A. | Counterpart Decl. | FHL Association | Counterpart Agreement?[a] |
|---|---|---|---|---|---|
| 4FGL J0221.5+2513 | fsrq | 02:21:26.97 | 25:14:33.67 | 3FHL J0221.4+2512 | No |
| 4FGL J0233.0+3740 | bcu | 02:33:7.99 | 37:41:59.83 | 3FHL J0233.0+3742 | Yes, J0233.0+3742 − 1 |
| 4FGL J0319.4−7045 | bcu | 03:20:9.21 | −70:45:33.59 | 3FHL J0319.2−7045 | Yes, J0319.2−7045 − 1 |
| 4FGL J0402.9+6433 | bcu | 04:02:54.45 | 64:35:10.08 | 3FHL J0402.9+6433 | Yes |
| 4FGL J0459.4+1921 | bcu | 04:59:31.50 | 19:22:41.99 | 3FHL J0459.3+1921 | No |
| 4FGL J0501.0+2424 | bcu | 05:01:6.92 | 24:23:18.12 | 3FHL J0501.0+2425 | Yes |
| 4FGL J0838.5+4013 | ... | ... | ... | 3FHL J0838.5+4006 | ... |
| 4FGL J0901.5+6711 | bcu | 09:01:40.80 | 67:11:58.48 | 3FHL J0901.5+6712 | No |
| 4FGL J1127.8+3618 | fsrq | 11:27:58.871 | 36:20:28.35 | 3FHL J1127.8+3615 | Yes, J1127.8+3615 − 1 |
| 4FGL J1421.4−1655 | ... | ... | ... | 3FHL J1421.5−1654 | ... |
| 4FGL J1626.0−4917c | ... | ... | ... | 3FHL J1626.3−4915 | ... |
| 4FGL J1729.9−4148 | ... | ... | ... | 3FHL J1729.9−4148 | ... |
| 4FGL J1808.8+2419 | bcu | 18:08:45.69 | 24:19:5.70 | 3FHL J1808.7+2420 | Yes |
| 4FGL J1849.3−6447 | bcu | 18:49:24.80 | −64:49:33.48 | 3FHL J1849.3−6448 | Yes |
| 4FGL J2026.6+3449 | bcu | 20:26:38.113 | 34:50:24.50 | 3FHL J2026.7+3449 | Yes |
| 4FGL J2109.6+3954 | ... | ... | ... | 3FHL J2109.6+3954 | ... |
| 4FGL J2317.7+2839 | ... | ... | ... | 3FHL J2317.8+2839 | ... |
| 4FGL J2358.5−1808 | bll | 23:58:36.84 | −18:07:17.49 | 3FHL J2358.4−1808 | Yes |

**Notes.** Table 4 outlines the 4FGL counterparts and their corresponding 3FHL associations as reported in the 4FGL-DR2. In a few cases, our X-ray detected counterpart position disagrees with the one associated in the 4FGL.
[a] Denotes if the 4FGL counterpart and our identified counterpart are spatially coincident. If there was more than one X-ray counterpart detected, the one that is coincident is listed.

blazar. Therefore, we consider SWIFT J032009−704535 to be the correct X-ray counterpart for 3FHL J0319.2−7045.

### 3.1.3. 3FHL J0233.0+3742

In this case, SWIFT J023308+374158 (see Figure 6) lies within both the 3FHL and 4FGL 95% uncertainty regions. The 3FHL object is associated with 4FGL J0233.0+3740, which has NVSS J023308+374201 as its associated counterpart. SWIFT J023308+374158 is spatially consistent with NVSS J023308+374201. Finally, SWIFT J023308+374158 has WISE IR colors that are compatible with a blazar. Alternatively, SWIFT J023256+373843 lies just outside the 4FGL 95% positional uncertainty region.

Even so, SWIFT J023256+373843 meets all of the criteria to be considered a blazar candidate. SWIFT J023256+373843 has WISE IR colors that place the source in the BL Lac region of the WISE blazar strip, it exhibits a power-law behavior at X-ray energies, and it is spatially coincident with the radio source VLASS1 J023256.08+373846.3. Because both X-ray sources meet our criteria to be a blazar candidate, both were kept as candidates. If both sources emitted gamma-rays, they would appear as a confused (single) source, due to the large LAT PSF. Only modeling their SEDs would reveal this potential case of source confusion.

### 3.1.4. 3FHL J1626.3−4915

SWIFT J162609−491746 lies within both the 3FHL and 4FGL 95% positional uncertainty regions, while SWIFT J162703−491230 lies outside (see Figure 7). Both X-ray sources have similar S/N and lack radio counterparts. However, SWIFT J162609−491746 is the only source, between the two, with WISE IR colors. In this case, the WISE IR colors align more closely with those exhibited by Galactic sources. The likely Galactic nature of this source is discussed in Section 3.4. SWIFT J162609−491746 is identified as the X-ray counterpart for 3FHL J1626.3−4915.

### 3.1.5. 3FHL J1127.8+3615

As Figure 8 shows, the 3FHL and 4FGL uncertainty regions have very little overlap. SWIFT J112759+362033, lies about 1.1′ from the 4FGL 95% positional uncertainty region, while SWIFT J112741+362051 lies approximately 1.7′ outside of the region. Again, the rather large LAT PSF makes it difficult to discern if either or both of these sources are emitting the gamma-rays being detected. SWIFT J112741+362051 has no WISE source consistent with its location, and hence no WISE colors to be compatible with those exhibited by blazars. 3FHL J1127.8+3615 is associated with 4FGL J1127.8+3618, which is associated to MG2 J112758+3620, a known FSRQ. We find SWIFT J112759+362033 to be coincident with MG2 J112758+3620, as well as with a WISE source that has colors consistent with the region overlapping BL Lacs and FSRQs. This makes SWIFT J112759+362033 our most likely X-ray counterpart.

## 3.2. 4FGL-DR2 Associations

The second release of the 4FGL (4FGL-DR2; Abdollahi et al. 2020; Ballet et al. 2020) provides detections of sources in 16 bands from 100 MeV to 1 TeV over 10 yr of exposure. It also provides updated associations for 12 of our 3FHL objects. Of these 12, nine are in agreement with the X-ray source discovered in the field (see Table 4). On the other hand, our favored X-ray counterpart is in disagreement with the association provided by the 4FGL for three sources. We discuss these in detail in the following sections.

### 3.2.1. 3FHL J0221.4+2512/4FGL J0221.5+2513

For 3FHL J0221.4+2512, we identify the counterpart to be SWIFT J022135+251418, while the 4FGL-DR2 counterpart is associated with 2MASS J02212698+2514338. The two contending counterparts are separated by a distance of ∼2′. We find that the Swift−XRT detected counterpart location

Table 5
X-ray Counterparts Spatially Coincident with Previously Identified Blazars

| XRT Name[a] | Catalog Name | Distance (″) | Blazar Type | Associated Paper |
| --- | --- | --- | --- | --- |
| J023308+374158 | 87GB 023007.1+372939 | 3.42 | BLL | D'Abrusco et al. (2019) |
| J112759+362033 | ICRF J112758.8+362028 | 4.92 | QSO | Massaro et al. (2009) |
| J235837−180718 | 6dFGS gJ235836.8−180717 | 1.32 | BLL | Lefaucheur & Pita (2017) |
| J083903+401547 | 2MASX J08390306+4015457 | 2.46 | BLL | Massaro et al. (2009) |

**Notes.** Table 5 outlines external manuscripts that identify our X-ray detection location as a blazar in another catalog.
[a] Sources detected with Swift−XRT referenced in this paper are prefaced with the SWIFT designation (e.g., SWIFT J023308+374158).

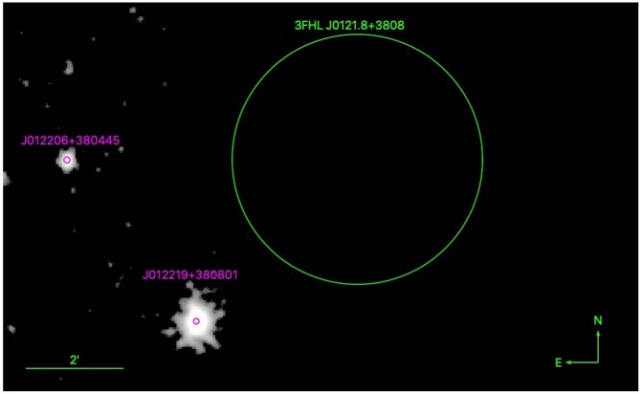

**Figure 4.** Image field for 3FHL J0121.8+3808 in the 0.3−10 keV energy range. No significant detections were determined to fall within the 3FHL uncertainty region for 3FHL J0121.8+3808. The XRT uncertainty regions are denoted in purple. J012206+380445 (upper source) and J012219+380801 (lower source) comprise our two counterpart candidates. A previous gamma-ray burst detection led to this field having greater counts on the eastern side of the field, possibly requiring additional time to distinguish a source within the 3FHL region.

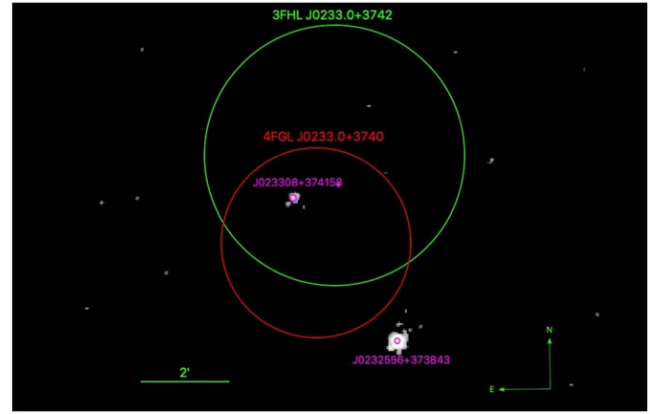

**Figure 6.** Swift−XRT image of 3FHL J0233.0+3742 field. In purple, we show SWIFT J023308+374158 (center) and SWIFT J023256+373843 (bottom right) with corresponding uncertainty regions. The 3FHL and 4FGL 95% uncertainty regions are plotted in green and red, respectively. Image energy range is 0.3−10.0 keV.

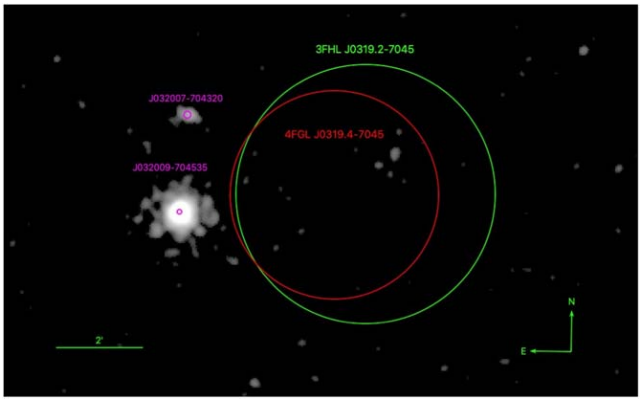

**Figure 5.** Multiple X-ray counterpart candidates for 3FHL J0319.2-7045 imaged in 0.3−10 keV. SWIFT J032009-704535 (bottom) and SWIFT J032007-704320 (top) are labeled with corresponding circular uncertainty regions. SWIFT J032009-704535 lies ~1.3′ and SWIFT J032007-704320 ~1.5′ from the edge of the 3FHL region. Multiwavelength data help to confirm that the brighter SWIFT J032009-704535 detection is most likely our corresponding X-ray counterpart.

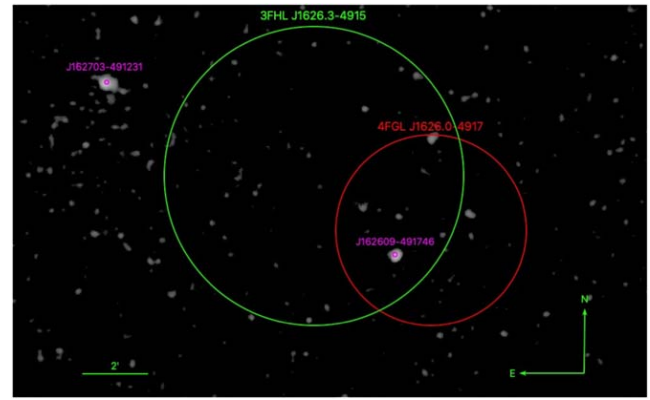

**Figure 7.** Swift−XRT image of 3FHL J1626.3-4915 field taken in 0.3−10 keV energy range. SWIFT J162703-491231 (upper left) and SWIFT J162609-491746 (lower right) both constitute X-ray counterparts for this field. Their corresponding pointing uncertainty regions (in purple) are plotted along with the 95% uncertainty regions for the 3FHL and 4FGL.

*3.2.2. 3FHL J0459.3+1921/4FGL J0459.4+1921*

lacks multiwavelength characteristics similar to blazars (see Table 6). On the other hand, 2MASS J02212698+2514338 has available data in radio, infrared, and optical emission, with its WISE IR colors being consistent with those of blazars. Therefore, we believe 2MASS J02212698+2514338 is the correct counterpart candidate to 3FHL J0221.4+2512, as it exhibits more blazar-like qualities.

This source is classified as a BCU in the 3FHL catalog. This source is associated with 1RXS J045931.5+192242. Our X-ray source, SWIFT J045928+192213, has flux compatible with the ROSAT source and is thus the likely counterpart. Indeed, 1RXS J045931.5+192242 has an unabsorbed 0.3−10.0 keV flux of $1.24 \pm 0.50 \times 10^{-12}$ erg cm$^{-2}$ s$^{-1}$ compared to $1.93 \pm 0.39 \times 10^{-12}$ erg cm$^{-2}$ s$^{-1}$ from Swift−XRT. Moreover, SWIFT J045928+192213's uncertainty region includes both a radio and a WISE source with IR colors that are indicative of a blazar. This strengthens the association to a blazar (see Table 7). Therefore, SWIFT J045928+192213 is selected as the most likely counterpart for 3FHL J0459.3+1921.

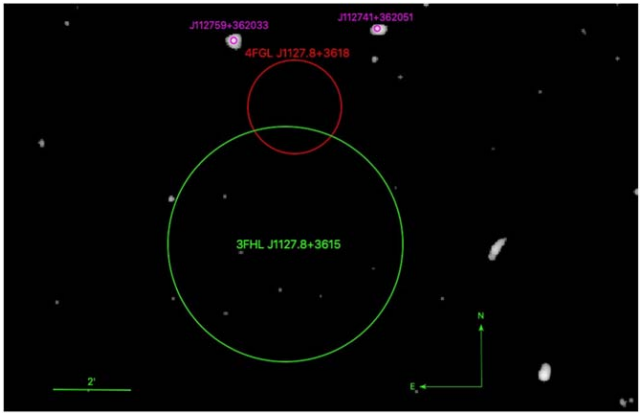

**Figure 8.** Swift−XRT image of 3FHL J1127.8+3615 in 0.3−10 keV energy range. Although the 4FGL was able to significantly reduce the 95% uncertainty region for this gamma-ray detection, both X-ray counterpart candidates fall outside of this region. Both SWIFT J112759+362033 (left) and SWIFT J112741+362051 (right) make persuasive blazar candidates. Due to the vicinity of the 4FGL uncertainty region, J112759+362033 was selected as the counterpart candidate.

### 3.2.3. 3FHL J0901.5+6712 / 4FGL J0901.5+6711

This 4FGL-DR2 source is associated with the ROSAT X-ray source 1RXS J090140.8+671158. Our XRT source, SWIFT J090135+671318, is the brightest source in the field. The two X-ray detections have an angular separation of over 80″ with no overlap in their 95% uncertainty regions. The 1RXS J090140.8+671158 calculated 0.3−10.0 keV flux is $1.03 \pm 0.44 \times 10^{-12}$ erg cm$^{-2}$ s$^{-1}$, compared to $0.66 \pm 0.15 \times 10^{-12}$. Furthermore, SWIFT J090135+671318 is positionally consistent with a radio source (see Table 8). This makes SWIFT J090135+671318 the most likely counterpart to 3FHL J0901.5+6712.

### 3.3. Chance Coincidence

We evaluate the chance of observing a background source within the 95% confidence region of a 3FHL object. We follow the procedure of Xi et al. (2020) to compute the chance probability for such a random coincidence as

$$P_{\rm ch} = 1 - \exp[-\pi(R_0^2 + 4\sigma_\gamma^2)\Sigma(>F_{\rm th})], \quad (1)$$

where $R_0$ is the median distance between the X-ray and gamma-ray detection, $\sigma_\gamma$ is the average 95% positional uncertainty radius of the 3FHL region, and $\Sigma(>F_{\rm th})$ is the X-ray source density for detections above the threshold flux, $F_{\rm th}$.

Relying on the X-ray 2–10 keV log$N$ − log$S$ of AGN (Ueda et al. 2014) we obtain a density of background AGN of 6 AGN per degree$^2$. This yields $P_{\rm ch} = 0.197$. This method, however, does not take into account the different exposures of the fields used here. Therefore, we used the 4XMM−DR11 (Webb et al. 2020) to calculate $\Sigma(>F_{\rm th})$, which resulted in a density of five X-ray sources per degree$^2$ (for the average 0.2–12 keV flux as measured in our fields). See Appendix B for details on this calculation. Using the corresponding value for the 3FHL 95% uncertainty region, the average uncertainty radius was $\sigma_\gamma = 0°.050$. The median distance from the X-ray detection to the gamma-ray detection was $R_0 = 0°.039$. Using these values resulted in a $P_{\rm ch} = 0.166$.

Utilizing our calculated $P_{\rm ch} = 0.166$, a probability mass function, which denotes the probability of a random discrete value being equivalent to a given value, was used to calculate the likelihood of detecting five fields (out of 21) with a background object (e.g., two or more X-ray sources). We find that the chance probability that five of our 21 fields would contain a double X-ray source detection is 14%. For perspective, for three fields, it is 23% (see Figure 9).

If we sum the results of Kaur et al. (2019), Silver et al. (2020), and this work, we find that 10 of the 98 fields (∼10%) observed contain a double X-ray detection. The mean value of the most likely number of fields predicted by $P_{\rm ch}$ is 16 fields. With a standard deviation of four fields, we find that our aggregate result is only 1.5$\sigma$ away from our predicted value. This tension is not statistically significant.

### 3.4. Low-latitude Sources

The 3FHL reports 133 extragalactic associations located at $|b| < 10°$, while 83 sources at the same latitudes are unassociated. The 3FHL authors estimate ∼25–40 of the 83 unassociated sources will be of Galactic origin, following the trend of the associated sources (Ajello et al. 2017). Of our 21 3FHL objects analyzed, five are low-latitude. All but one of them exhibit the characteristics typical of blazars.

The distribution of Galactic sources in the 3FHL peaks at $|b| < 2°$, with all other latitudes reporting substantially fewer sources (see Figure 11 in Ajello et al. 2017). Located at $|b| = 0°.11$, 3FHL J1626.3−4915 was identified as having the X-ray counterpart SWIFT J162609−491746. 3FHL J1626.3 −4915 is associated with 4FGL J1626.0−4917c (the "c" denotes the use of caution in analyzing or interpreting this source, as the source is confused with interstellar cloud complexes), with no counterpart or class association reported in the catalog. The WISE IR colors coincident with our counterpart's position fall in line with other classified Galactic sources (unassociated low-latitude source in Figure 3, far left). Specifically, the WISE W2 magnitude is greater than W1, a common characteristic of Galactic sources. Due to these reasons, we consider 3FHL J1626.3−4915 a Galactic source. Based on its gamma-ray photon index of 2.5 ± 0.42, in comparison to other sources in the 3FHL it is most indicative of a pulsar wind nebulae (PWN) or a supernova remnant (Ajello et al. 2017). Further investigation is necessary to confirm the classification of this source.

## 4. Classification with Machine Learning

Machine-learning algorithms can enable us to achieve our goal of predicting the source classification (FSRQ or BL Lac) for our blazar candidates. We implemented three machine-learning algorithms (decision trees (Breiman 2001), random forests (Breiman 2001), and support vector machines (Zhou 2021)) to create source classification models for our 21 blazar candidates. Each algorithm was trained by a data set. The data set consisted of 438 known 3FHL BL Lac and FSRQ blazars whose corresponding 4FGL reported counterparts are spatially coincident with sources in the 2SXPS (Evans et al. 2020) and AllWISE (Cutri et al. 2021) catalogs. The six features (gamma-ray photon index, X-ray photon index, variability Bayesian blocks, and WISE IR colors W1−W2,

Table 6
4FGL J0221.5+2513/SWIFT J022135+251418 Multiwavelength Comparison

| 4FGL Counterpart | Unc. (″) | XRT Name[a] | Unc. (″) |
|---|---|---|---|
| 4FGL J0221.5+2513 | 0.001 | J022135+251418 | 4.33 |
| Band | Counterpart | Band | Counterpart |
| Radio | NVSS 022126+251436 | Radio | ... |
| Infrared | AllWISE J022126.96+251433.6 | Infrared | ... |
| Optical/UV | GAIA 1029337016124175336 | Optical/UV | ... |
| X-ray | ... | X-ray | J022135+251418 |

**Notes.** Multiwavelength comparison of the 4FGL (left) and XRT (right) counterpart candidates in the field for 3FHL J0221.4+2512.
[a] Sources detected with Swift−XRT referenced in this paper are prefaced with the SWIFT designation (e.g., SWIFT J023308+374158).

W2−W3, and W3−W4) used for determining classification were selected because they have been observed to best differentiate BL Lacs and FSRQs. BL Lacs typically exhibit a harder gamma-ray spectrum (Abdo et al. 2010; Ackermann et al. 2015) and a softer X-ray spectrum than FSRQs (Donato et al. 2001), leading us to select the X-ray and gamma-ray indices from the 2SXPS and 3FHL catalog (Ajello et al. 2017), respectively. Variability Bayesian Blocks (VBB) is a feature provided in the 3FHL catalog that lists the number of Bayesian blocks detected when the source changed state. VBB ranges from 1 to 15, where 1 implies no variability and 15 implies highest variability. In general, FSRQs are more variable than BL Lacs and thus have higher VBB values. We have already shown how WISE IR color–color space can distinguish BL Lacs and FSRQs into distinct regions (see Figure 3). Table 9 reports the complete list of training features and their respective references.

### 4.1. Decision Tree

A decision tree classifier is a supervised machine-learning algorithm that separates a data set into distinct branches and nodes based on a set of features. This separation repeats itself until each data point is a distinct node at the end of a continuous set of branches. The DT will continue to branch until the probability of a random source being incorrectly labeled (also known as the impurity) is minimized (Breiman 2001). Each of these nodes is assigned one of the categorizations based on the training set. The results of a single run of the DT on our 21 blazar candidates are given in Table 10.

### 4.2. Random Forest

A random forest classifier acts as an aggregation of DT iterations (Breiman 2001). Since it utilizes multiple decision trees, the problem of overfitting is eliminated (which is usually observed in the case of a single decision tree; see Gu 2014). A final classification is given based on the ensemble results. Moreover, this method reports the likelihood probabilities for sources classified as BL Lacs and FSRQs. These values are specified in Table 10.

### 4.3. Support Vector Machines (SVM)

Another supervised learning algorithm used in this study is the support vector machine. SVMs work on the intrinsic principle that, for two distinguishable data sets, one or more sets of maximum margin hyperplanes can be established such that the distance from the plane to the closest point in either category is maximized (Zhou 2021). We employed a polynomial kernel and a nonlinear SVM to make our classifications. The probabilities associated with each classification along with the resulting classifications are reported in Table 10.

### 4.4. Machine-learning Implementation

We employed the algorithms of DT, RF, and SVM, as provided by `sklearn 1.0.2` package in Python 3.1, to create classification models that operated on our 21 uncategorized sources from the 3FHL.

Of the 21 blazar candidates, four lacked WISE data. For these sources, the mean and median was calculated for the aggregate WISE colors from our 17 remaining blazar candidates. The four sources' classifications were unchanged by using the mean or the median.

Feature importance for a random forest algorithm is evaluated by the aggregate mean and standard deviation of the impurity decrease within each tree. Therefore, features that decrease the probability of a random source being incorrectly classified are more heavily favored (Breiman 2001). Unlike Kaur et al. (2019) and Silver et al. (2020), in this work, W3−W4 was a feature element included in the models. As seen in Figure 10, this IR color is more significant for classifying sources than other features used previously in Kaur et al. (2019) and Silver et al. (2020). Because of this significance, the W3−W4 color was included in our final classification.

The accuracy of a machine-learning model can be measured by testing the model on a subcategory of the training set that was not used to train the model (typically referred to as a validation set; Meyer-Baese & Schmid 2014). Because the classifications of the validation set are known, when the model is applied to a validation set, the probability of a correct classification is denoted as the true-positive rate (TPR). On the other hand, the false-positive rate (FPR) is the probability of a spurious classification. A receiver operating characteristic (ROC) curve plots these two rates, with the upper left corner indicating a better classifier. The diagonal from the lower left to the upper right indicates where a no-skill or a chance-based algorithm would fall on the plot. The accuracy is determined by measuring the area under the ROC curve (Metz 1978). Any source classified above the no skill value (50%) sets our minimum threshold for the cutoff in classification.

### 4.5. Results

The DT model resulted in 19 sources being classified as BL Lacs, while one was classified as an FSRQ. The DT model

Table 7
4FGL J0459.4+1921/SWIFT J045928+192213 Multiwavelength Comparison

| 4FGL Counterpart | Unc. (″) | XRT Name[a] | Unc. (″) |
|---|---|---|---|
| 1RXS J045931.5+192242 | 32.4 | J045928+192213 | 3.90 |
| **Band** | **Counterpart** | **Band** | **Counterpart** |
| Radio | ... | Radio | VLASS1QLCIR J045927.50+192215.3 |
| Infrared | ... | Infrared | AllWISE J045927.49+192215.0 |
| Optical/UV | ... | Optical/UV | GAIA 3408317206647436672 |
| X-ray | 1RXS J045931.5+192242 | X-ray | J045928+192213 |

**Notes.** Table 7 displays the multiwavelength data comparison of the 4FGL (left) and XRT (right) counterpart candidates in the field for 3FHL J0459.4+1921.
[a] Sources detected with Swift−XRT referenced in this paper are prefaced with the SWIFT designation (e.g., SWIFT J023308+374158).

yielded an accuracy of 85% when assessed with the validation set. The RF and SVM models predicted 13 out of 21 blazar candidates to be identified as BL Lacs with a >90% classification probability. The seven remaining sources were identified as BL Lacs at lower likelihood probabilities. These results coincide with our initial intuition that the unassociated sources of the 3FHL should reflect the prevalence of BL Lacs in the associated sources. The ROC for the RF model reported an accuracy of 95% (see Figure 11). For the SVM model, the ROC yielded an accuracy of 93% (see Figure 12). The number of points present in the ROC is established by the algorithm used.

## 5. Discussion and Conclusion

This project continues the campaign to identify and classify the remaining 200 unassociated and unknown sources in the 3FHL catalog. Developing an accurate machine-learning algorithm could enable reliable predictions for classifying the remaining unassociated and unknown sources, eliminating the need for multiple nights of optical observation. The prevalence of blazars in the 3FHL motivates our hypothesis that most unassociated sources in the 3FHL are also blazars. Blazars are scientifically relevant because they provide constraints for high-energy astrophysics. Completing the catalog would provide additional constraints on the gamma-ray horizon (Ajello et al. 2015; Ackermann et al. 2016), as well as expand the number of blazars used to measure the extragalactic background light (EBL; LAT Collaboration 2018).

Blazars comprise 95% of all 3FHL classified sources that share the range of gamma-ray photon index and flux (in the 10 GeV−1 TeV band) spanned by the 21 unassociated and unknown sources analyzed by this work. We regard this as evidence that most of our sources are likely blazars. Additional multiwavelength data further helped to inform us which sources are most likely blazar counterparts.

From a combination of proposed and archival sources, 30 unassociated 3FHL sources with Swift−XRT data were analyzed. Of these fields, 16 had a single X-ray source detected in the field, five fields had more than one X-ray source present, and seven had no source detected in a typical ∼2 ks observation (with two fields having the X-ray source ruled out as a likely star). This resulted in 26 X-ray source detections. On the basis of their multiwavelength properties, 21 X-ray sources were designated as candidate blazars and one source was designated to be of Galactic origin. The seven fields where no X-ray source was detected have a mean exposure time of ∼5 ks. This lack of detection could be caused by the source being intrinsically faint or at high redshift. We recommend future Swift−XRT observations of these sources in order to identify the appropriate counterpart. The list of all sources analyzed as a part of this work are reported in Appendix C.

Machine-learning models were trained using spectral properties of known blazars. The random forest and support vector machine models developed to test our observed sources predicted that all 21 blazar candidates would be classified as BL Lacs. Collectively, Kaur et al. (2019), Silver et al. (2020), and this work have predicted the classification of 72 sources. The RF and SVM models (when applicable) predicted all 72 sources to be BL Lacs. Discrepancies arise in the DT model, which is prone to weaknesses such as overfitting (Gu 2014). The lack of any FSRQ classifications by the RF and SVM methods across all three works supports the ubiquity of BLLs in the 3FHL. On the other hand, the lack of a single FSRQ being classified for 72 unassociated sources could indicate that our models are biased in favor of classifying BL Lacs, thus leading to the misclassification of some FSRQ blazars.

As Table 4 indicates, based on 4FGL-DR2 classifications, we would have expected two of our sources, 3FHL J0221.4+2512 and 3FHL J1127.8+3615, to be classified as FSRQs. The release of the 4FGL-DR3 (Data Release 3; LAT Collaboration et al. 2022) occurred concurrently with this work. The 4FGL-DR3 provides the most up-to-date associations and classifications for the gamma-ray sky. In this release, 4FGL J0221.5+2513 (which is associated with 3FHL J0221.4+2512) is classified as a BL Lac instead of as an FSRQ, its classification in the 4FGL-DR2. Both releases report the same counterpart, 2MASS J02212698+2514338, which is in spatial agreement with our reported X-ray counterpart.

On the other hand, our predicted class and observed classification of 3FHL J1127.8+3615 still disagree. Although the counterpart identified in this investigation, SWIFT J112759+362033, and the counterpart reported in both data releases of the 4FGL, MG2 J112758+3620, are spatially coincident, the optical observations at the location of this counterpart (SDSS J112758.86+362028.3) confirm the source to be an FSRQ. This misclassification is likely due to the WISE counterpart colors falling in the overlapping BL Lac and FSRQ regions of the WISE blazar strip, thus making it difficult to distinguish the classification with these features.

We express caution with regard to the model predictions of 3FHL J0500.6+1903, 3FHL J0901.5+6712, and 3FHL J2026.7+3449. These sources lack WISE color data, so the average WISE color of all counterparts analyzed was used (as described in Section 4.4). Thus, these values will be intrinsically biased toward BL Lacs. Utilizing multiwavelength data in classifying blazars should give more accurate results,

Table 8
4FGL J0901.5+6711/SWIFT J090135+671318 Multiwavelength Comparison

| 4FGL Counterpart | Unc. (″) | XRT Name[a] | Unc. (″) |
|---|---|---|---|
| 1RXS J090140.8+671158 | 32.4 | J090135+671318 | 4.46 |
| Band | Counterpart | Band | Counterpart |
| Radio | ... | Radio | NVSS J090133+671317 |
| Infrared | ... | Infrared | ... |
| Optical/UV | ... | Optical/UV | GALEX UV J090134.1+671316 |
| X-ray | 1RXS J090140.8+671158 | X-ray | J090135+671318 |

**Notes.** Table 8 identifies the multiwavelength data comparison of the 4FGL (left) and XRT (right) counterpart candidates in the field for 3FHL J0901.5+6711.
[a] Sources detected with Swift−XRT referenced in this paper are prefaced with the SWIFT designation (e.g., SWIFT J023308+374158).

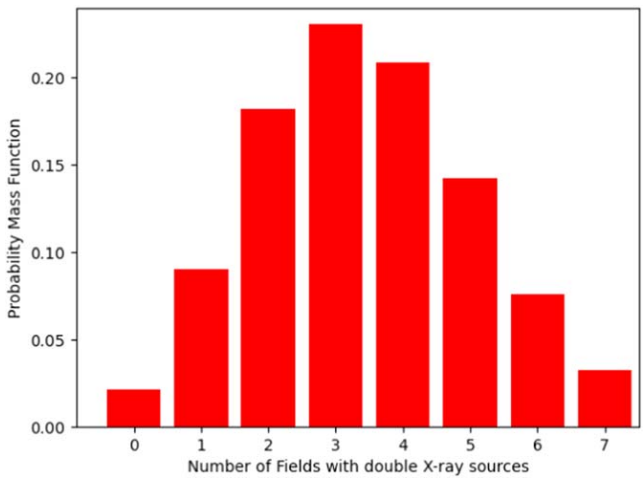

**Figure 9.** The probability mass function plotted for the likelihood to detect a double X-ray detection in a given number of fields out of the 21 fields observed. The probability of detecting five fields with a double X-ray detection is 14%.

Table 9
Features for Blazar Classification

| Feature | Catalog | Reference |
|---|---|---|
| Gamma-ray Photon Index | 3FHL | Ajello et al. (2017) |
| X-ray Photon Index | Table 1 for unknown sample | See Table 1 |
|  | 2SXPS for training set | Evans et al. (2020) |
| W1−W2 | AllWISE | Cutri et al. (2021) |
| W2−W3 | AllWISE | Cutri et al. (2021) |
| W3−W4 | AllWISE | Cutri et al. (2021) |
| Variability Bayesian Blocks | 3FHL | Ajello et al. (2017) |

**Note.** This table indicates the features used by our machine-learning models to classify our sources.

although it may be more difficult to obtain data for all sources in all bands. These reasons motivate us to improve our machine-learning classification model by other methods.

Numerous other works have used machine-learning techniques in a variety of ways to classify possible AGN. Parkinson et al. (2016) utilized a random forest to predict if gamma-ray sources in the 3FGL were likely to be young pulsars, millisecond pulsars, or AGN. Building upon the likely AGN as predicted in Parkinson et al. (2016), Salvetti et al. (2017) exclusively used gamma-ray flaring properties to optimize an artificial neural network algorithm to classify 456 3FGL unassociated sources as BL Lacs or FSRQs. For the 27 sources that were both classified by Salvetti et al. (2017) and observed with optical spectroscopy, their model predictions were ∼90% accurate. Although gamma-ray properties play a pivotal role in determining classification, we believe the inclusion of multiwavelength data will strengthen the accuracy of classification models.

Besides the works directly related to this campaign, Kerby et al. (2021a) is an example of using X-ray counterpart spectral data to develop a classification model that is trained with multiwavelength data (i.e., gamma-ray and X-ray). Just as we used WISE data to create multiple new features, one can combine other multiwavelength properties into an original feature to be used for a machine-learning algorithm. Kerby et al. (2021a) used the logarithmic ratio between X-ray and gamma-ray flux and found this to be a significant feature, which was used in a subsequent paper (Kerby et al. 2021b). Kerby et al. (2021b) also notes that neural networks tend to perform better than random forest algorithms, with increased confidence and a reduction in the number of ambiguous sources.

The accuracy of our models was determined by testing them with a test set of blazars with known classifications. For the sources studied in this work, our predicted classifications will by verified (or refuted) when the sources are observed by optical spectroscopy follow-ups. In the future, it may be advantageous to implement features from other works that have been successful in accurately classifying AGN and blazars. Given that our sources are not notably variable, other significant features from Parkinson et al. (2016) shared in the 3FHL such as the curvature (Signif_Curve) and uncertainty in flux above 100 MeV (Unc_Energy_Flux100) may increase our model's accuracy. A future investigation that examines which features give the most accurate classifications may be a worthwhile endeavor. A more robust model could even possibly unveil additional blazar types when applied en masse to known blazars.

Finally, the lack of redshifts as reported in the 3FHL for all 21 of our unassociated and unknown sources fuels a greater need for spectroscopic analysis. Utilizing the optical/UV data procured from Swift−UVOT, we plan to have optical follow-up observations to measure the spectroscopic redshift of our sources. Both Sbarufatti et al. (2006) and Álvarez Crespo et al. (2016) have shown that small, 4 m telescopes are not productive for measuring redshift for BL Lacs. Previous optical surveys done as a part of this campaign reflect this as

Table 10
Machine-learning Categorization Results

| 3FHL Name | DT Class | RF Class | RF Probability | SVM | SVM Probs | 4FGL Counterpart Classification[a] |
|---|---|---|---|---|---|---|
| 3FHL J0121.8+3808 − 1[b] | BLL | BLL | 0.86 | BLL | 0.78 | ... |
| 3FHL J0221.4+2512 | BLL | BLL | 0.98 | BLL | 0.99 | fsrq |
| 3FHL J0233.0+3742 − 1[b] | BLL | BLL | 0.91 | BLL | 0.99 | bcu |
| 3FHL J0233.0+3742 − 2[b] | BLL | BLL | 0.86 | BLL | 0.76 | bcu |
| 3FHL J0319.2−7045 −1[b] | BLL | BLL | 1.00 | BLL | 0.99 | bcu |
| 3FHL J0402.9+6433 | FSRQ | BLL | 0.68 | BLL | 0.89 | bcu |
| 3FHL J0459.3+1921 | BLL | BLL | 0.92 | BLL | 0.80 | bcu |
| 3FHL J0500.6+1903 | BLL | BLL | 1.00 | BLL | 0.99 | ... |
| 3FHL J0501.0+2425 | BLL | BLL | 0.88 | BLL | 0.78 | bcu |
| 3FHL J0838.5+4006 | BLL | BLL | 1.00 | BLL | 1.00 | ... |
| 3FHL J0901.5+6712 | BLL | BLL | 0.88 | BLL | 0.92 | bcu |
| 3FHL J0950.6+6357 | BLL | BLL | 0.84 | BLL | 0.81 | ... |
| 3FHL J1127.8+3615 − 1[b] | BLL | BLL | 0.79 | BLL | 0.88 | fsrq |
| 3FHL J1421.5−1654 | BLL | BLL | 1.00 | BLL | 0.94 | ... |
| 3FHL J1729.9−4148 | BLL | BLL | 1.00 | BLL | 0.99 | ... |
| 3FHL J1808.7+2420 | BLL | BLL | 0.86 | BLL | 0.91 | bcu |
| 3FHL J1849.3−6448 | BLL | BLL | 0.97 | BLL | 0.82 | bcu |
| 3FHL J2026.7+3449 | BLL | BLL | 1.00 | BLL | 0.97 | bcu |
| 3FHL J2109.6+3954 | BLL | BLL | 0.69 | BLL | 0.54 | ... |
| 3FHL J2317.8+2839 | BLL | BLL | 0.62 | BLL | 0.63 | ... |
| 3FHL J2358.4−1808 | BLL | BLL | 1.00 | BLL | 0.97 | bll |

**Notes.** Table 10 displays the machine-learning results for the three models used to determine blazar type for blazar candidates: decision tree (left), random forest (center), and support vector machine (right). Included on the far right is the classification as reported by the 4FGL-DR2.
[a] The 4FGL denotes an association with lowercase letters.
[b] The additional digit was used to distinguish sources where more than one was present in a single field.

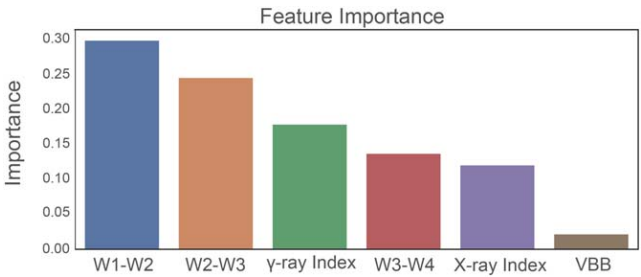

**Figure 10.** Random forest feature importance for class identification. The addition of the W3–W4 feature was included because it is more significant than two other features used in previous campaigns of this kind (see Kaur et al. 2019; Silver et al. 2020).

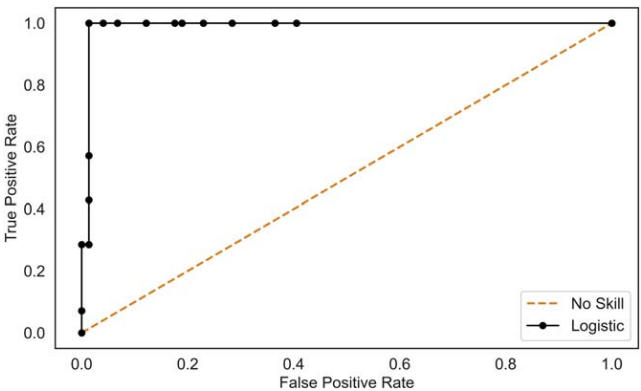

**Figure 11.** ROC curve from the random forest method on the validation sample. The area under the curve yielded 0.95. The diagonal line represents the nondiscriminatory curve. Hence, any point below this line would not be classified by the model.

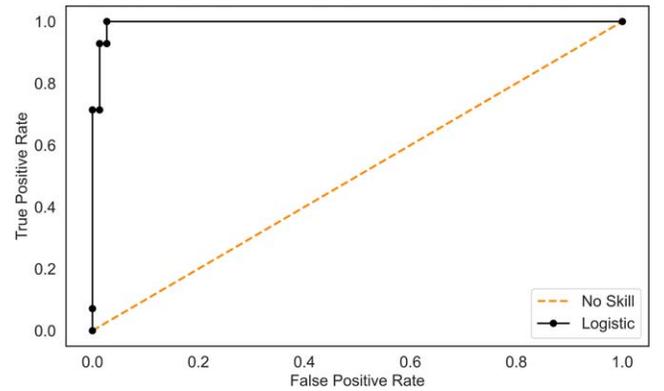

**Figure 12.** ROC curve resulting in an accuracy of 0.93. This curve comes from the SVM model used on our 21 blazar counterpart candidates. The effectiveness of the classifier is measured by the area under the curve, with 1 being the ideal value i.e., all true positives and no false positives.

well (Marchesi et al. 2018; Desai et al. 2019; Rajagopal et al. 2021). Of these three sets of observations, Rajagopal et al. (2021) was the most successful, placing redshift constraints for 57% of the observed sources. Although 4 m class telescopes are ineffectual for measuring redshift for BL Lacs, each campaign successfully classified the sources observed. In order to measure redshift effectively, larger 8 m or 10 m observatories are required (Paiano et al. 2017a, 2019). By obtaining spectroscopic data from our sources, we will be able to verify the source classifications from our models. This next step will bring us closer to realizing the completion of the 3FHL catalog.

We would like to thank the reviewer, data editor, and statistics expert who reviewed our paper for their insight and recommendations. We graciously acknowledge NASA funding under grant 80NSSC21K2079, which enabled this work to be done. This research has made use of the VizieR catalog access tool, CDS. Additionally, this research has made use of the SIMBAD database, operated at CDS, Strasbourg, France. This research has made use of data and/or software provided by the High Energy Astrophysics Science Archive Research Center (HEASARC), which is a service of the Astrophysics Science Division at NASA/GSFC and the High Energy Astrophysics Division of the Smithsonian Astrophysical Observatory. This publication made use of data products from the Wide-field Infrared Survey Explorer, which is a joint project of the University of California, Los Angeles, and the Jet Propulsion Laboratory/California Institute of Technology, funded by the National Aeronautics and Space Administration. This publication makes use of data products from the Two Micron All Sky Survey, which is a joint project of the University of Massachusetts and the Infrared Processing and Analysis Center/California Institute of Technology, funded by the National Aeronautics and Space Administration and the National Science Foundation. This work has made use of data from the European Space Agency (ESA) mission Gaia (https://www.cosmos.esa.int/gaia), processed by the Gaia Data Processing and Analysis Consortium (DPAC, https://www.cosmos.esa.int/web/gaia/dpac/consortium). Funding for the DPAC has been provided by national institutions, in particular the institutions participating in the Gaia Multilateral Agreement. Finally, we thank the reader for their interest in our work.

*Software:* Astropy (The Astropy Collaboration et al. 2013), FTOOLS (Blackburn 1995), HEASoft (v6.281; NASA High Energy Astrophysics Science Archive Research Center (HEASARC), 2014), numpy (Harris et al. 2020), SAO Image DS9 (Joye & Mandel 2003), scikitlearn (Pedregosa et al. 2011), TOPCAT (Taylor 2005), XSPEC (v12.11.1; Arnaud 1996).

## Appendix A
### Catalogs Used for Multiwavelength Characterization

SIMBAD, Vizier, and CDS Xmatch were used to find multiwavelength detections that are spatially coincident with our XRT detections. Of the catalogs present in these databases, the ones specifically used for this project are listed here.

1. Radio: NVSS, VLASS1, ILT, and PMN.
2. Infrared: AllWISE, 2MASS.
3. UV/Optical: GALEX, GAIA EDR3, and SDSS.
4. Additional Galaxy catalogs: 87GB, ICRF, 6dFGS, and 2MASX.

The appropriate data for 2MASS and AllWISE can be found through IPAC at Skrutskie et al. (2006) and Wright et al. (2019), respectively.

## Appendix B
### Coincidence Calculations

$F_{th}$ was calculated from modeling an absorbed power law using the median column density ($N_H = 0.057$ cm$^{-2}$) of our X-ray sources detected by Swift between 0.3–10.0 keV. We adapted the parameters until they met the median flux from our X-ray detections (Flux = $3.5034 \times 10^{-13}$ erg cm$^{-2}$ s$^{-1}$). Extrapolating the model to XMM-Newton's energy range (0.2–12.0 keV), a flux of $3.7572 \times 10^{-12}$ erg cm$^{-2}$ s$^{-1}$ is the minimum threshold flux $F_{th}$ that would correspond to a similar source. The EPIC and PN camera's full-range band was used to find all sources that met this $F_{th}$. This resulted in 6153 sources per 1239 degree$^2$, or ~5 sources per degree$^2$.

## Appendix C
### Sources Analyzed

Here, we list the 30 3FHL sources that were observed with Swift-XRT and analyzed as part of this investigation.

1. 3FHL J0110.9+4346
2. 3FHL J0121.8+3808
3. 3FHL J0221.4+2512
4. 3FHL J0233.0+3742
5. 3FHL J0243.3+1915
6. 3FHL J0301.4-5618
7. 3FHL J0319.2-7045
8. 3FHL J0402.9+6433
9. 3FHL J0459.3+1921
10. 3FHL J0500.6+1903
11. 3FHL J0501.0+2425
12. 3FHL J0550.4-4356
13. 3FHL J0550.9+5657
14. 3FHL J0753.9+0452
15. 3FHL J0838.5+4006
16. 3FHL J0901.5+6712
17. 3FHL J0950.6+6357
18. 3FHL J1127.8+3615
19. 3FHL J1421.5-1654
20. 3FHL J1528.4-6730
21. 3FHL J1602.8-1928
22. 3FHL J1626.3-4915
23. 3FHL J1729.9-4148
24. 3FHL J1808.7+2420
25. 3FHL J1849.3-6448
26. 3FHL J2026.7+3449
27. 3FHL J2109.6+3954
28. 3FHL J2245.5-1734
29. 3FHL J2317.8+2839
30. 3FHL J2358.4-1808.

ORCID iDs

S. Joffre https://orcid.org/0000-0001-9427-2944
R. Silver https://orcid.org/0000-0001-6564-0517
M. Rajagopal https://orcid.org/0000-0002-8979-5254
M. Ajello https://orcid.org/0000-0002-6584-1703
N. Torres-Albà https://orcid.org/0000-0003-3638-8943
A. Pizzetti https://orcid.org/0000-0001-6412-2312
S. Marchesi https://orcid.org/0000-0001-5544-0749
A. Kaur https://orcid.org/0000-0002-0878-1193